\documentclass[pre,amsmath,twocolumn,amssymb,showpacs]{revtex4}
\usepackage{epsfig}
\usepackage{graphicx}
\usepackage{dcolumn}
\usepackage{bm}
\usepackage[english]{babel}

\def\Ref#1{(\ref{#1})}

\begin{document}

\title{Shear induced crystallization of an amorphous system}

\author{Anatolii V. Mokshin}
\affiliation{Universit\'e de Lyon; Univ. Lyon
I,  Laboratoire de Physique de la Mati\`ere Condens\'ee et des
Nanostructures; CNRS, UMR 5586, 43 Bvd. du 11 Nov. 1918, 69622
Villeurbanne Cedex, France }

\author{Jean-Louis Barrat}
\affiliation{Universit\'e de Lyon; Univ. Lyon
I,  Laboratoire de Physique de la Mati\`ere Condens\'ee et des
Nanostructures; CNRS, UMR 5586, 43 Bvd. du 11 Nov. 1918, 69622
Villeurbanne Cedex, France }

\date{\today}

\begin{abstract}
The influence of a stationary shear flow on the crystallization in
a glassy system is studied by means of molecular dynamics
simulations and subsequent cluster analysis. The  results reveal
two opposite effects  of the shear flow  on the processes of
topological ordering in the  system. Shear promotes the formation
of separated crystallites and suppresses the appearance of the
large clusters. The shear-induced ordering proceeds  in two
stages, where the first stage is related mainly with the growth of
crystallites, whereas the second stage is due to an adjustment of
the created clusters and  a progressive alignment of their lattice
directions. The influence of strain and shear rate on the
crystallization is also investigated. In particular, we find two
plausible phenomenological relations between the shear rate and
the characteristic time scale needed for ordering of the amorphous
system under shear.
\end{abstract}

\maketitle

\section{Introduction}

Most liquids, under cooling, undergo a first order transition to a
crystalline phase. The classical view is that this transition
takes place through an homogeneous nucleation process and can be
reasonably well described in the framework of classical nucleation
theory \cite{Debenedetti,Kelton,Barrat,Chaikin,Lebowitz,Sear}.
Nucleation theory is based on the fact, that a nucleation event is
an activated process, taking place on   time scales much  larger
than the characteristic time scale of the microscopic dynamics.
The free energy of forming a crystalline embryos from the
metastable surroundings is defined by a positive surface and a
negative bulk contributions. The surface term corresponds to the
cost in free energy for creation of an interface between parent
and incipient (say, crystalline) phases, whereas the bulk term is
proportional to the volume of the nucleus. The crystal size,
wherein the free energy reaches a maximum and the system begins to
crystallize, defines the critical nucleus.

With the increase of the degree of supercooling $\Delta T$ the
description of the transition towards the ordered phase becomes
more complicated. On the one hand, the height of nucleation
barrier decreases with supercooling $\Delta T$ as $1/\Delta T^2$,
so that at supercooling $\sim 40\%$ and higher a very fast crystal
nucleation could be expected \cite{Wolde}. On the other hand,
observations at a very large supercooling indicate  nascent
droplets that exhibit a ramified structure \cite{Yang} and a
crystallization process with a more extended, collective and
spatially scattered character that may be attributed to a spinodal
regime \cite{Trudu}. Such tendencies are, however, balanced by the
kinetic slowing down, which makes the ordering process more and
more difficult to observe as the temperature is lowered. In the
limit, where the system is deeply supercooled and becomes glassy,
crystal formation is completely unobservable on experimental time
scales.

The application of external field  on a glassy material may change
considerably this picture of nucleation. What influence has an
external forcing, such as shear flow and/or strain, on the
nucleation process of a glass? Experimental studies of amorphous
(co)polymers reveal the appearance of shear-induced
crystallization \cite{Pluta,Naudy,Duplay,Lellinger}. This is
verified by molecular dynamics simulations, which provide some
evidence for an increase of ordering in a sheared polymeric and
model binary glasses \cite{Wallace,Albano/Falk}. Recently, results
of molecular dynamics simulations showed that the
\textit{oscillating shear strain} can promote crystallization in a
model jammed systems \cite{Duff}. However, the influence of shear
rate and strain on the ordering processes  as well as the
possibility of crystallization under stationary shear have
remained unclear.

At a moderate supercooling, simulation results on the sheared
colloidal melts of Refs. \cite{Blaak1,Blaak2} demonstrate the
suppression of nucleation by a  homogeneous shear flow. More
precisely, they reveal that the probability of the nucleation
decreases, while the size of critical nuclei increases with the
shear rate. In a glassy systems at low temperatures, on the other
hand, it is reasonable to suggest that the  external drive can
\textit{activate} dynamical processes \cite{Ilg/Barrat,Haxton}. A
moderate external shear field will increase the local diffusivity
in the system, hence having a positive influence on the kinetic
factors for nucleation, and thereby it will trigger
crystallization. However, a continued shear may destroy a
crystalline nuclei as they form and  a steady nonequilibrium state
can be expected.

In the present work we focus on the influence of a
\textit{stationary shear} on the ordering processes in a  glassy
system. In Sec. \ref{system} we describe our model system and the
analysis used to identify crystallinity and solid-like clusters.
The simulation results and outcomes of cluster analysis are shown
in Sec. \ref{results}, where we also study the  influence of
strain and shear rate on the ordering. We finish in Sec.
\ref{nature} with a discussion of the main results.

\section{System and procedures} \label{system}

Our system consists of $23328$ particles interacting through a
standard truncated and shifted Lennard-Jones potential
\begin{eqnarray}
U(r)= \left\{
\begin{array}{lc}  \displaystyle
4\epsilon \left [  \left ( \frac{\sigma}{r} \right )  ^{12}  -
\left ( \frac{\sigma}{r} \right ) ^{6} \right . \\
\ \ \ \ \ \ \ \ \ \ \   \displaystyle \left .  - \left (
\frac{\sigma}{r_c} \right ) ^{12} + \left ( \frac{\sigma}{r_c}
\right ) ^{6} \right ], &
 r \leq  r_c \\
0, & r > r_c,
\end{array}
\right. \label{hevisaid}
\end{eqnarray}
$\epsilon$ and $\sigma$ are the characteristic energy and length
scales, and $r_c=2.5\sigma$ is the cutoff distance. The following
reduced units are used in this work. The time is in units of
$\tau=\sigma \sqrt{m/\epsilon}$, where the mass $m$ is set to unity.
All distances are given in units of $\sigma$. The temperature is in
units of $\epsilon/k_B$, whereas the pressure and the stress are in
units of $\epsilon/\sigma^3$. The time step  $\triangle \tau$ used
in our simulations is $0.005\tau$.

We start with a  system equilibrated in the liquid state at the
temperature $T=1.65\epsilon/k_B$ in a cubic simulation box ($V=L^3$
and $L=30.2327\sigma$) with periodic boundary conditions in all
directions. After this, the system is quenched to the temperature
$T=0.15\epsilon/k_B$ within a time interval  $t=2.5\tau$. In the
case of argon atoms with the Lennard-Jones parameters $\epsilon/k_B
= 120$~K and $\sigma=3.4$\AA~ this corresponds to a cooling rate
$\sim 10^{13}$~K/s \cite{Rahman}. After such a rapid quench, the
system at this temperature is in a glassy state
\cite{Mazzacurati,Ruocco,Santis,Simdyankin}. This is evidenced by
the zero slope of mean square displacement on the time scale of our
simulations, which is a signature of  ``structural arrest''. The
disordered character of the structure is also evident from the split
second peak of the pair distribution function, typical of an
amorphous material.

\begin{figure}[tbh]
\centerline{\psfig{figure=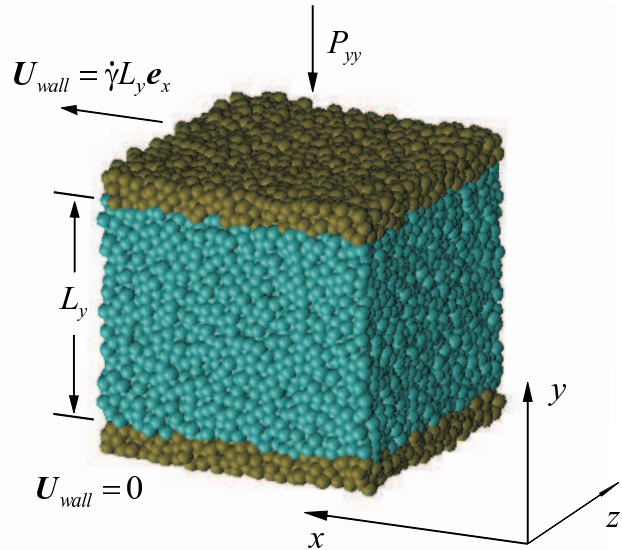,height=8cm,angle=0}} \caption{
(Color online) A snapshot of the simulation cell. Two parallel
amorphous walls (dark particles) restrict the  sheared system. The
particles of the top wall are removed with the velocity proportional
to the distance between the walls $L_y$, which is variable with the
time due to the normal pressure $P_{yy}$. The particles of the
bottom wall are fixed.} \label{snapshot}
\end{figure}

Before shearing, the system is allowed to "age"  during  $t =
10~000$~$\tau$ without any external forcing. In order to shear the
system, we create two parallel solid walls by freezing all the
particles in the $x$-$z$ plane over the range of three
inter-particle distances from both ends of the simulation box in
the $y$-direction. Both walls are amorphous.  By using walls, we
can  impose an average strain rate without any assumptions about
the resulting flow inside of the sample. A snapshot of the
simulation cell is presented in Fig. \ref{snapshot}. Twelve
independent samples were prepared with the same procedure.

A constant shear rate $\dot{\gamma}$ is then applied by moving in
the $x$-direction all the atoms of the top wall with the
instantaneous velocity
\begin{equation}
\textit{\textbf{U}}_{wall} = \dot{\gamma}L_y
\textit{\textbf{e}}_x, \label{vel_top}
\end{equation}
whereas the particles of the bottom wall remain fixed; $L_y$ is the
distance between the walls. Periodic boundary conditions are applied
along the $x$ and $z$ directions only. All the  results are for a
constant normal pressure $P_{yy}=1.1867\epsilon/\sigma^3$
(corresponding to the pressure observed initially). Temperature is
controlled by rescaling of velocity component of the  particles
along the neutral $z$-direction, which is perpendicular to the shear
$x$ and the velocity gradient $y$ directions.

In order to identify the local structure and, in particular, the
appearance of solid-like clusters, we use the local order analysis
introduced originally in the work of Steinhardt \textit{et al.}
\cite{Steinhardt} and developed  by Frenkel and co-workers
\cite{Wolde,Sanz,Butler}. An important advantage of this method is
that (i) it allows one to recognize the crystallinity regardless of
a specific structure, and (ii) the crucial parameters here, such as
local and global order parameters, are rotationally invariant, so
that the orientation of clusters in  space is irrelevant.

First of all, the local surroundings of each atom can be
characterized by a ($2\times6+1$)-dimensional complex vector with
the following components:
\begin{equation}
q_{6m}(i) =   \frac{1}{N_b(i)} \sum_{j=1}^{N_b(i)}
Y_{6m}(\theta_{ij}, \varphi_{ij} ), \label{q_6_var}
\end{equation}
where $Y_{6m}(\theta_{ij}, \varphi_{ij})$ are the spherical
harmonics, $N_b(i)$ denotes the number of the nearest neighbors of
particle $i$; $\theta_{ij}$ and $\varphi_{ij}$ are the polar and
azimuthal angles formed by the radius-vector $\textbf{r}_{ij}$ and
some reference system. We define ``neighbors'' as all atoms
located within a given radius $r_c=1.5\sigma$ around an atom $i$,
i.e. $|\textbf{r}_{ij}|<r_c$, where $r_c$ corresponds practically
to the first minimum in the pair distribution function. The local
orientational order parameter can be defined for each atom $i$ as
\begin{equation}
q_6(i)=\left ( \frac{4\pi}{13} \sum_{m=-6}^{6}  | q_{6m}(i) |^2
\right )^{1/2},
\label{local_order_par}
\end{equation}
which is a rotationally invariant. Thus, the global orientational
order parameter can be defined as an average of $q_{6m}(i)$ over
all $N$ particles:
\begin{equation}
\mathcal{Q}_6 = \left ( \frac{4\pi}{13} \sum_{m=-6}^{6} \left |
\frac{\sum_{i=1}^{N} \sum_{j=1}^{N_b(i)} Y_{6m}(\theta_{ij},
\varphi_{ij})}{\sum_{i=1}^{N} N_b(i)} \right |^2 \right )^{1/2}.
\label{global_order_par}
\end{equation}
For perfect fcc, bcc and hcp systems one has $\mathcal{Q}_6 =
0.5745$ and $\mathcal{Q}_6 = 0.5106$ and $\mathcal{Q}_6 = 0.4848$,
respectively, whereas in a fully disordered system in the limit of
large sizes,  (e.g. a liquid) $\mathcal{Q}_6$ is close to zero
\cite{Rintoul}. Hence, an increase of this quantity provides the
evidence for the formation of local crystallites. To estimate the
degree of ordering (crystallinity) in our system we use, along with
$\mathcal{Q}_6$, the potential energy as well as the number of
``solid-like'' particles.

The occurrence of  ordered structures can  also be  observed in the
behavior of  the radial distribution function. However, this
function corresponds to  the averaged result for the whole system,
and, as a consequence of this, can be insensitive to the appearance
of a few local clusters.

For the study of  local structures we apply the following cluster
analysis \cite{Wolde}. For every pair of  nearest neighbors, say $i$
and $j$, the following condition is considered:
\begin{equation}
\left | \sum_{m=-6}^{6} \widetilde{q}_{6m}(i)
\widetilde{q}_{6m}^{*}(j) \right | > 0.5, \label{cond_coherence}
\end{equation}
where $\widetilde{q}_{6m}(i)$ is the complex vector $q_{6m}(i)$
defined by Eq. \Ref{q_6_var} and normalized in accordance with
\begin{equation}
\sum_{m=-6}^{6} \widetilde{q}_{6m}(i)\widetilde{q}_{6m}^*(i) = 1.
\end{equation}
Condition \Ref{cond_coherence} allows one to verify that   atom
$j$ belongs to an ordered structure around  atom $i$. If  atom $i$
has seven or more \textit{neighbors} satisfying the condition
\Ref{cond_coherence}, then this atom is considered as a
solid-like, i.e. it is included in an ordered crystalline
structure.

\section{Results} \label{results}

\begin{figure}[tbh]
\centerline{\psfig{figure=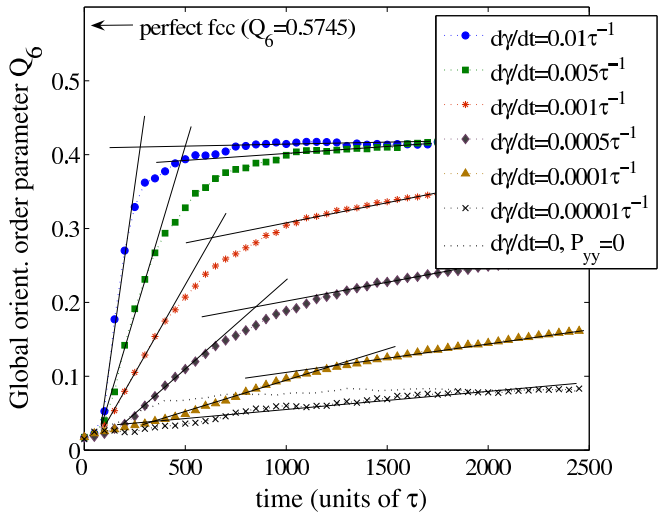,height=7.5cm,angle=0}}
\caption{ (Color online) The evolution of the global orientational
order parameter $\mathcal{Q}_6$ at the various shear rates
$\dot\gamma$ as a function of time.  The shear rate increases from
bottom to top. The full lines are linear fits to the data. The
arrow indicates the value of $\mathcal{Q}_6$ for a perfect fcc
structure. } \label{Q_6_time}
\end{figure}

We now turn to the results, which quantify the effect of shear on an
initially amorphous sample. The time evolution of the global
orientational order parameter $\mathcal{Q}_6$ for  various values of
the shear rate $\dot{\gamma}$  is shown  in Fig. \ref{Q_6_time}.
Note that the results presented here have been averaged over
different runs.
From these results, it appears clearly that the degree of order in
the system is enhanced by  shear. Moreover, at the larger values
of the shear rate $\dot{\gamma}$ the ordering takes place rather
rapidly,  $\mathcal{Q}_6$ levels off after this initial transient.
For $\dot{\gamma}=0.01$ and $0.005$$\tau^{-1}$ the order parameter
reaches a plateau value with $\mathcal{Q}_6 \approx 0.42$ over the
time scale of observation. At lower values of $\dot{\gamma}$
\textit{the rate of ordering} (defined as the time derivative of
the order parameter) is lower and decreases with the time, as it
is clear seen from the  change in slope in the curves shown in
Fig. \ref{Q_6_time}. As a result,   the time window $t=2500\tau$
presented in Fig. \ref{Q_6_time} is not sufficient to achieve a
maximal ordering at slow shear rates. It is particularly
noteworthy that the shear can initially prevent the formation of
small clusters, which would  appear even in the absence of shear.
This week suppression effect is observed for very small shear
rates $\dot{\gamma}=0.0001$ and $0.000~01$~$\tau^{-1}$, where the
values of $\mathcal{Q}_6$ are lower in comparison with the case of
a sample at rest. Nevertheless, the increase of the order
parameter with time is clearly detected even for these small shear
rates. Our first conclusion is therefore that  shear  enhances
 crystallinity with a rate, which depends on the shear rate.

\begin{figure}[tbh]
\centerline{\psfig{figure=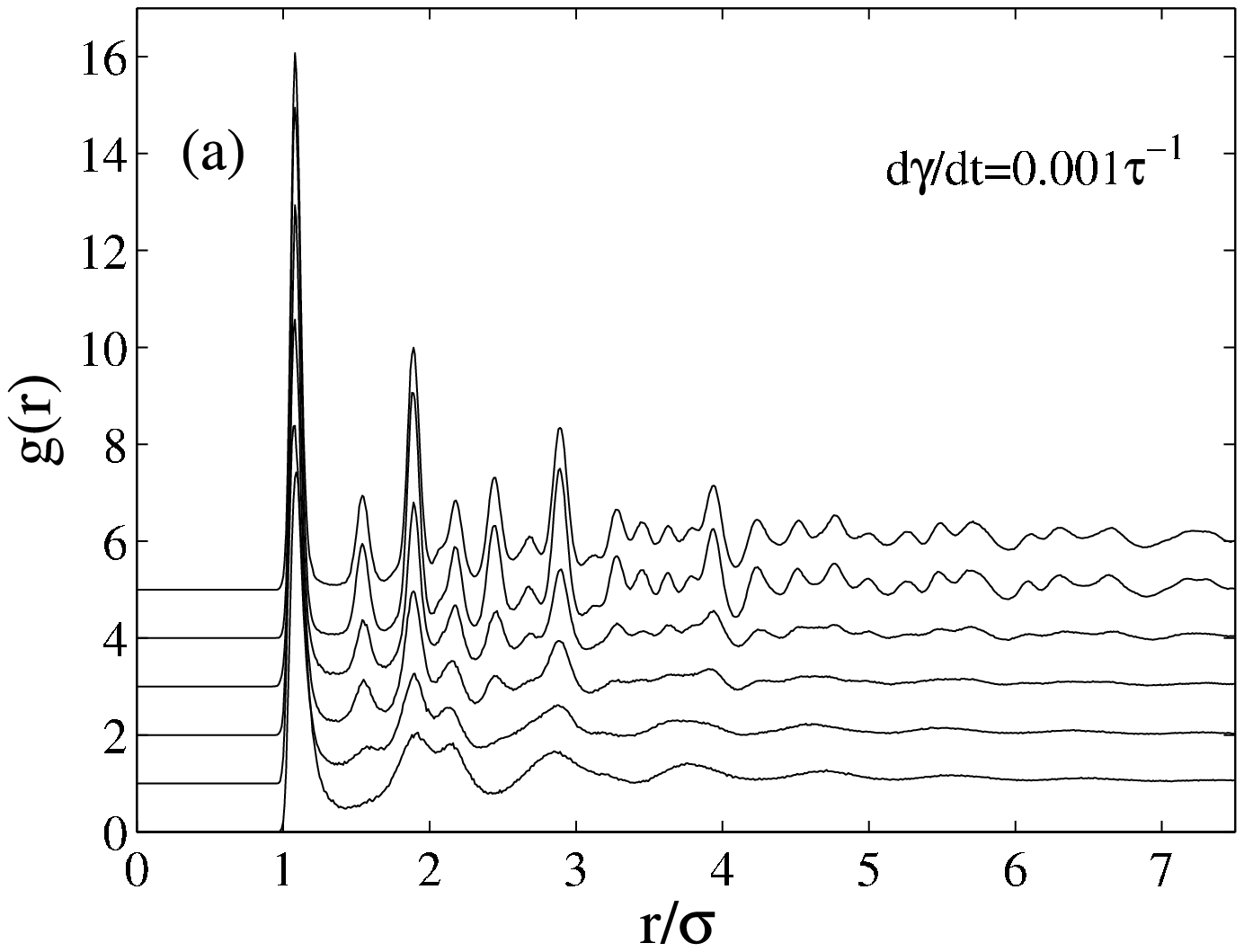,height=6.5cm,angle=0}}
\centerline{\psfig{figure=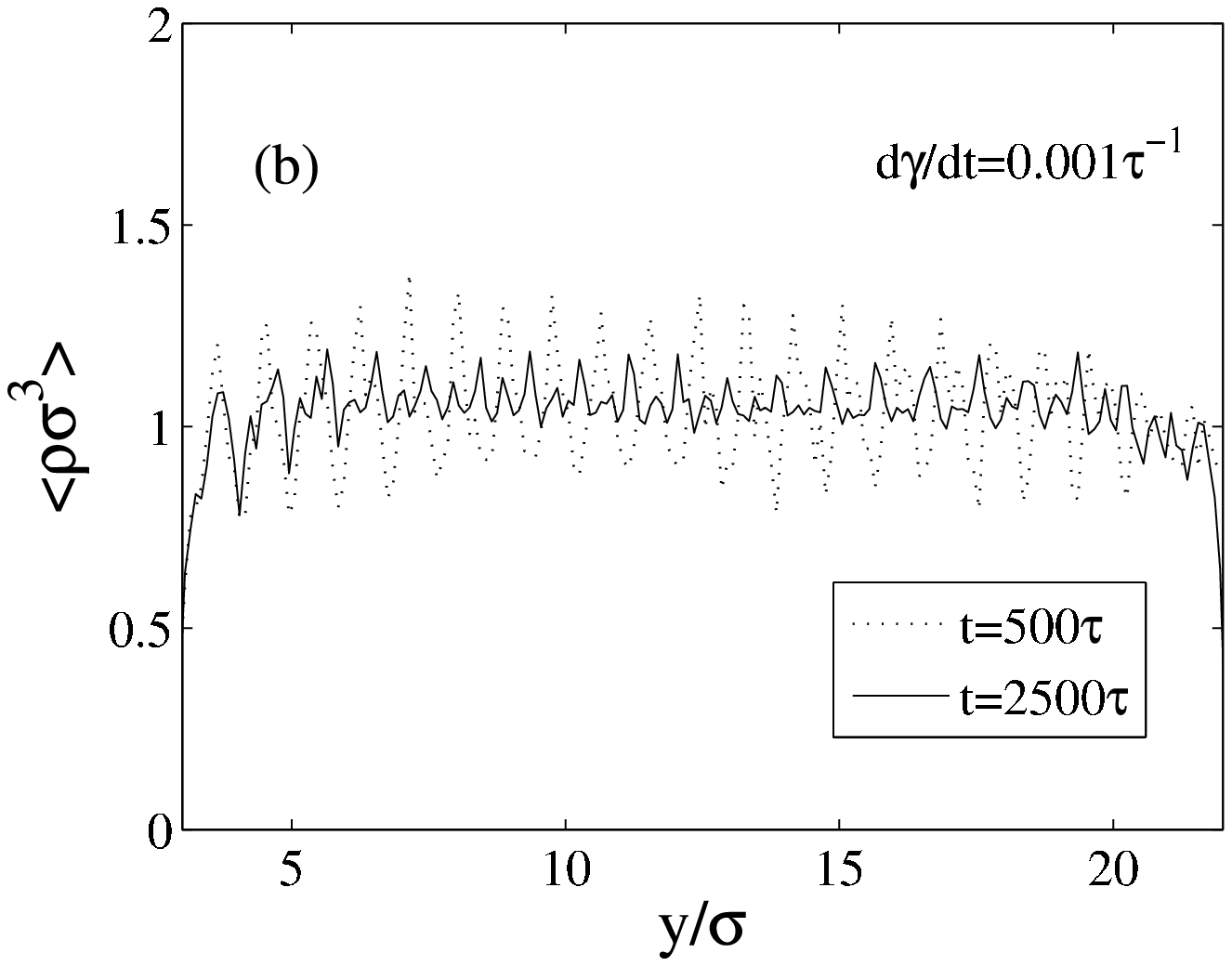,height=6.5cm,angle=0}}
\caption{Structural characteristics of the sample under shear (at
the shear rate $\dot\gamma = 0.001$$\tau^{-1}$): (a) Radial
distribution function at the different times after starting the
shear flow: $t=0$, $250$, $500$, $750$, $1500$ and $2500$$\tau$
(from bottom to top). The curves are shifted upwards for clarity.
(b) Density profiles for two different times as averaged over a
time window of $10\tau$. } \label{rdf}
\end{figure}

Although the largest value of the order parameter $\mathcal{Q}_6$
obtained by shearing the  glass  is low in comparison with
$\mathcal{Q}_6$ of a perfect fcc structure, it indicates a high
level of crystallinity in the  system. The formation of crystalline
ordered structures in a sample can be also observed from the radial
distribution function as shown in Fig. \ref{rdf}(a) for a particular
shear rate $\dot\gamma = 0.001$ and  different times after starting
up of the shear. In the first three curves the appearance of
crystalline structures is evident from  the rise of the extra-peak
between the first and second maxima in the distribution, which is a
typical signature of fcc structures. The pair correlation function
calculated  at large times shows that the order extends  over large
distances, with oscillations extending up  to  $r>4\sigma$. Such a
long-range ordering could also be associated  with   layering of the
system under shear. To check this, we evaluated the density profiles
at different times as a function of the distance from the walls. As
can be seen in Fig. \ref{rdf}(b), which presents the density
profiles at $t=500$ and $2500$$\tau$ (the curves for intermediate
times are very similar to those presented in Fig. \ref{rdf}(b)), the
transverse order is present, but not particularly pronounced.

Moreover, the layering is weaker for  $t=2500$$\tau$, whereas the
pair correlation function displays the more pronounced structure.
This observation leads to the conclusion that the  long-range order
observed in the pair correlations is   caused mainly by the
formation of a crystalline  clusters.

\subsection{Strain and strain rate dependance}

\begin{figure}[tbh]
\centerline{\psfig{figure=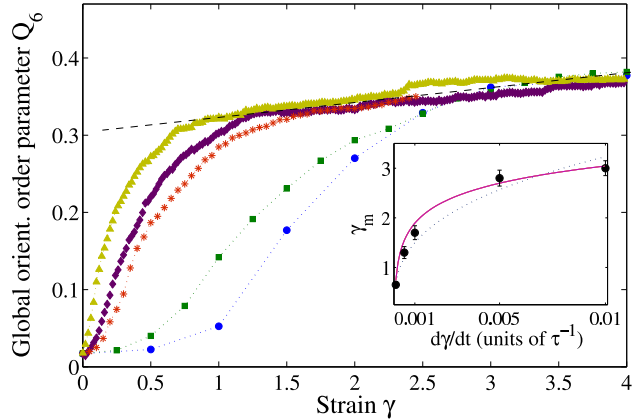,height=6cm,angle=0}}
\caption{(Color online) Main: Strain dependence of the global
orientational order parameter $\mathcal{Q}_6$ at the different shear
rates $\dot\gamma = 0.0001$, $0.0005$, $0.001$, $0.005$ and
$0.01$$\tau^{-1}$ (from left to right). The correspondence between
curves and shear rates is the same as in Fig. \ref{Q_6_time}. The
dashed line is the linear interpolation, indicating the steady
behavior (plateau) in the strain dependence of order parameter.
Inset: Shear rate dependence of the strain  $\gamma_m$, at which the
order parameter reaches  the plateau value. Errors are defined by
the change of slope in $\mathcal{Q}_6(\gamma)$. The solid and dotted
curves are the logarithmic and power-law approximations,
respectively (see text). } \label{strain}
\end{figure}

The strain dependence of the global order parameter $\mathcal{Q}_6$
is shown in Fig. \ref{strain}. It can be seen that the order
parameter increases with the strain $\gamma$ for all values of the
shear rate. Moreover, the evolution of the order  parameter with
strain is clear separated into two steps. The first stage, during
which the  the order parameter rises rapidly, is shear rate
dependent. After this fast increase, $\mathcal{Q}_6$ reaches an
``universal'' (shear rate independent) behavior shown  by a straight
dashed line in Fig. \ref{strain}. At this stage, the order parameter
demonstrates a very slow increase and eventually  levels off  at
large strains.

In order to quantify the influence of shear rate on the ordering, we
introduce the ``ordering strain'' $\gamma_m$. This quantity defines
the strain scale, where the evolution of $\mathcal{Q}_6$ merges with
the universal behavior shown by a dashed line in Fig. \ref{strain}.
The shear rate dependence of $\gamma_m$ is plotted in the inset of
Fig. \ref{strain}. The data for  $\gamma_m(\dot{\gamma})$ can be
fitted   either by a power-law or by a logarithmic dependency on the
strain rate:
\begin{subequations} \label{gamma_m}
\begin{equation}
\gamma_m \propto \dot{\gamma}^{1+n},\ n=-2/3,
\label{power_gamma_m}
\end{equation}
\begin{equation}
\gamma_m = \gamma_0 + \frac{1}{2} \ln(\dot{\gamma}), \ \ \gamma_0
= \textrm{const}. \label{ln_gamma_m}
\end{equation}
\end{subequations}
It should be noted, that  the power law behavior is supported by the
idea that the typical relaxation time $t_{\alpha}$ of a  sheared
glassy system decreases with the strain rate as
$\dot{\gamma}^{-2/3}$ (see Ref. \cite{Barrat_Kurchan}). Assuming
that a similar dependency holds in our crystallizing system and the
initial rise of $\mathcal{Q}_6$ corresponds to a typical relaxation
time, we obtain
\begin{equation}
t_m \sim \dot{\gamma}^{-2/3}. \label{JL_&_Kurchan}
\end{equation}
where $t_m= \gamma_m/\dot{\gamma}$. Such a  power-law decay of the
crystallization time with the shear rate could  also be related to
the  one found experimentally in Refs. \cite{Naudy,Lellinger}.

Although the possible relation between the characteristic time scale
$t_{\alpha}$ and the time of ordering $t_m$ under shear is
attractive \cite{Barrat_Kurchan}, it is seen in the inset of Fig.
\ref{strain} that the power-law with \Ref{power_gamma_m} does not
provide a perfect fit of $\gamma_m$ at  shear rates $\dot{\gamma}
\approx 0.01$$\tau^{-1}$ and higher, whereas the logarithmic
dependence \Ref{ln_gamma_m} gives a good fit to the data for all
values of the shear rate.

We finally discuss our results with regard to recent observations
reported by Duff and Lacks \cite{Duff}. These authors  studied the
ordering of a similar system under an \emph{oscillatory} strain in
the low temperature and  low shear rate limit. A degree of
ordering comparable to the one observed in our study was obtained
after two cycles with the amplitude 0.25. As a result, it would
correspond to a total strain   $\gamma_m=0.5$, which is comparable
to the values obtained by us at the lowest shear rates.

\subsection{Nature of the semi-crystalline state}

\begin{figure}[tbh]
\centerline{\psfig{figure=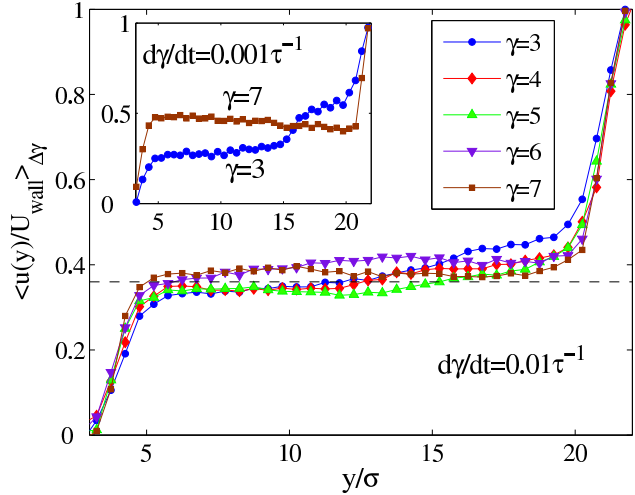,height=7cm,angle=0}}
\caption{(Color online) Rescaled velocity profile as a function of
distance from the bottom (unmoved) wall. Main: shear with
$\dot{\gamma}=U_{wall}/L_y = 0.01$$\tau^{-1}$ at different points of
the strain $\gamma$ above $\gamma_m = 3$. The broken line
corresponds to $u/U_{wall}= 0.36$. Inset:  shear rate $\dot{\gamma}
= 0.001$$\tau^{-1}$ at  two different strains above $\gamma_m$.
Results are averaged over the time window $t =
\Delta\gamma/\dot{\gamma}$, where $\Delta\gamma = 0.01$ is the
strain scale. All  runs exhibit a similar behavior.}
\label{vel_profile}
\end{figure}

\begin{figure}[tbh]
\centerline{\psfig{figure=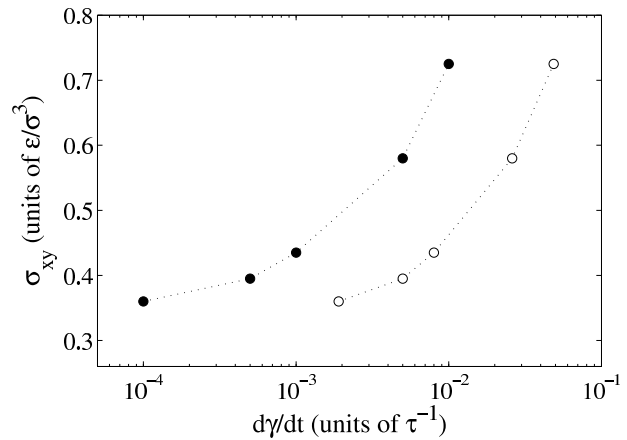,height=6.5cm,angle=0}}
\caption{ Connected empty circles: Shear stress versus strain rate
in the sheared part of the velocity profiles shown in Fig.
\ref{vel_profile}. Connected full circles: stress as a function of
the average shear rate in the sample. These curves are indicative
of a coexistence between a rapidly flowing shear band and a
non-flowing solid below its yield stress. } \label{stressstrain}
\end{figure}

Although the system clearly becomes more ordered under the influence
of strain, the degree of order achieved by our system is low in
comparison to that of a perfect crystal. A very remarkable fact is
that, at  large strains, the order parameter $\mathcal{Q}_6$ appears
to be dependent only on strain and not on strain rate. This
observation indicates some ``universality'' of the semi-crystalline
state created in the system. This observation is, however, easily
explained by considering   the velocity profiles in the sheared
systems. The velocity profiles presented in Fig. \ref{vel_profile}
exhibit a strong localization of the shear in two shear bands
located near the solid walls. The semi-crystalline part of the
sample, on the contrary, flows with an almost uniform velocity
(although some plastic activity is also taking place in this
``non-flowing'' part).
The situation described here is very similar to the  shear
localization observed in a flowing glass by Varnik \textit{et al.}
\cite{Varnik}. The nanocrystalline solid is submitted to a stress,
which is insufficient to cause flow, while two strongly fluidized
bands sustain  the shear entirely. What is remarkable here, this
is a high value of the shear rate and the stress, at which this
coexistence is observed (see Fig. \ref{stressstrain}).  While the
yield of the solid in Ref. \cite{Varnik} was observed for a strain
rate slightly above $0.001$$\tau^{-1}$ and a stress of
$0.6$$\epsilon/\sigma^3$, our results indicate here a yield stress
$\sigma_Y > 0.8$$\epsilon/\sigma^3$.

\begin{figure}[tbh]
\centerline{\psfig{figure=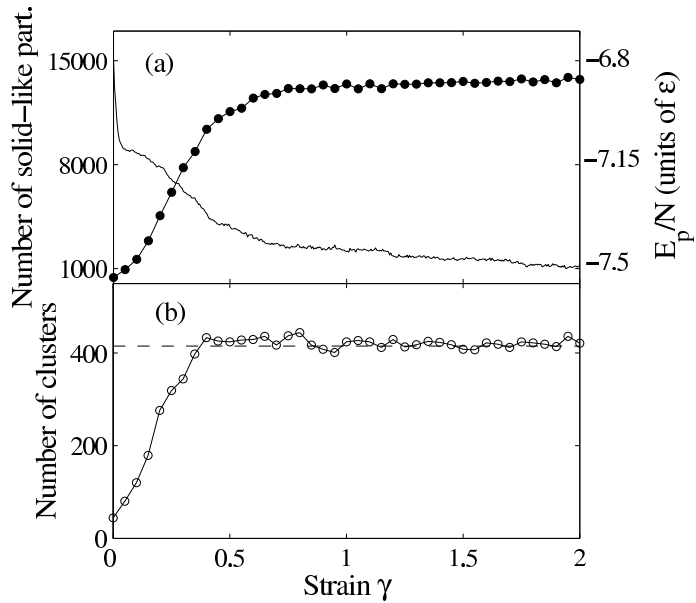,height=8.5cm,angle=0}}
\caption{ Different characteristics \textit{vs.} strain at the
shear rate $\dot{\gamma}=0.001$$\tau^{-1}$ after a single run: (a)
Number of solid-like particles in a whole system (connected full
circles) and potential energy per particle (solid line). (b)
Number of crystallites (connected open circles). The dashed line
corresponds to the value $415$.} \label{Pot_cluster}
\end{figure}

We expect that this nanocrystalline state consists in an assembly of
crystallites with disordered orientations. This set of crystallites
can be quantitatively described by means of the cluster analysis
presented in Sec. \ref{system}. The results of this analysis are
shown in Fig. \ref{Pot_cluster} as a function of the strain $\gamma$
for a particular shear rate, since the results for other values of
the shear rate are very similar.
The sample is imperfectly crystallized in such a  way that only
$\sim 80\%$ of the  particles are involved in  crystalline clusters.
Obviously, the rapid build up of crystalline order between
$\gamma=0$ and $\gamma_m$ corresponds to a rapid decrease in the
potential energy [see Fig. \ref{Pot_cluster}(a)]. The next
interesting feature is that the number of solid-like clusters [see
Fig. \ref{Pot_cluster}(b)], after a significant growth with the
strain, remains essentially constant, whereas the evolution of the
potential energy and of the order parameter indicates a continued
ordering process in the system. This leads to the conclusion  that
the system under shear decomposes rapidly into a set of crystalline
``grains''. The subsequent evolution consists in rearrangements
involving the grinding of grain boundaries and the alignment  of
neighboring grains without any significant coarsening.

\begin{figure}[tbh]
\centerline{\psfig{figure=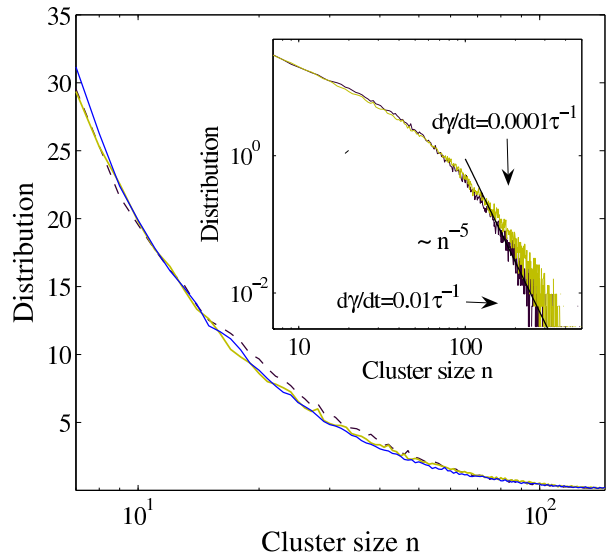,height=8cm,angle=0}}
\caption{(Color online) Main: Cluster size distribution at the
different shear rates $\dot{\gamma} = 0.01$, $0.001$ and
$0.0001$$\tau^{-1}$ (dashed, thin and thick lines, respectively) and
for a strain of $400\%$, where the system is characterized by the
order parameter $\mathcal{Q}_6 \approx 0.37$ for all shear rates
(see Fig. \ref{strain}). Each histogram is averaged over different
runs. Inset: The same distribution in log-log representation. The
solid line is the interpolation of the tail of the distribution at
$\dot{\gamma}=0.01$$\tau^{-1}$ by a power-law dependence.}
\label{distribution_cluster}
\end{figure}

Another interesting information about the final state of the
system can be extracted from the size distribution of the
crystalline grains, which is presented in Fig.
\ref{distribution_cluster} at three different shear rates. As can
be seen from these histograms, the distribution is dominated by
small grains involving less than $50$ particles. At first sight,
the distribution appears to be, within the accuracy of our data,
shear rate independent. This is consistent with the universal
behavior of the order parameter observed in Fig. \ref{strain}.
Nevertheless, the distribution displays a relatively slow decaying
tail for  cluster sizes larger than $100$ particles. This decay
can be well approximated by a power law (see inset of Fig.
\ref{distribution_cluster}). Closer examination reveals that the
weight of this tail depends weakly on shear rate and that larger
clusters can be observed at  lower shear rates. Unfortunately, the
small number of large clusters makes a systematic investigation of
this effect difficult.

\section{Discussion: shear suppression versus shear enhancement}  \label{nature}

Our results demonstrate  clearly that the shear increases initially
the tendency of a one-component amorphous system towards crystalline
order. This behavior should be discussed in the context of the
recent studies by Blaak and L\"owen, who demonstrated on the
contrary a shear suppression of the nucleation rate at moderate
undercooling \cite{Blaak1}, and of the recent results of Ref.
\cite{Trudu} concerning the evolution of the nucleation barrier with
temperature. Clearly, the main influence of shear at  low
temperature will be on the kinetic, rather than on the
thermodynamic, aspects of the transition. In the absence of shear,
the diffusivity is essentially zero, so that the system does not
evolve with time.  However, according to the classical nucleation
picture, one would expect the appearance of  crystallinity in the
form of a few isolated nuclei after a significant time lag
associated with the free energy barrier. Our results, on the
contrary, indicate an instantaneous increase of crystalline order as
soon as the shear is started, that is more consistent with a
spinodal description. The system is rapidly driven towards a new
energy minimum as soon as a mobility is reinstalled by the shear
flow. The order appears uniformly inside the system, which relaxes
locally towards a crystalline structure on a time scale, which is
characteristic of a sheared glass.

After this initial relaxation, a much slower stage of defect and
grain boundary annealing takes place. During this second stage, the
state of the system appears to be independent on the strain rate and
is determined by the amount of strain only. The system consists of
two rapidly flowing sheared bands, separated by a slab of a
nanocrystalline solid, which undergoes a very progressive evolution
through plastic rearrangements. This nanocrystalline solid appears
to have a  high yield stress in comparison with a similar
Lennard-Jones glass.

It is remarkable that, although the flow rate at the boundary
increases, the evolution of the solid slab seems to be insensitive
to this flow rate. A possible explanation is in the fact that the
energy in a yield stress system is dissipated by the flow, which
will serve to activate annealing processes in the solid slab. As a
result, the local structural transformations are defined essentially
by deformations and insensitive to strain rate.

The non-flowing part of the system can be mainly described as a
collection of the crystalline grains of relatively small size. The
stationarity in the number of crystallites indicates that the
disruption of crystalline order by the shear  at the boundaries
compensates completely the coarsening process, which would be
expected in a system with a nonzero atomic diffusion.

Finally, it appears that the shearing of an initially amorphous
one-component system  constitutes a reproducible way to obtain a
nanocrystalline state, which was sometimes taken in the past as a
possible model of an amorphous systems. It will be interesting to
study such a  state for its structural, vibrational and
rheological properties, that should be intermediate between those
of a glass and of a perfect crystal.

\begin{acknowledgments}

We thank A. Tanguy for useful discussions. Financial support from
ANR project SLLOCDYN is acknowledged. The simulations were
performed with the classical molecular dynamics package LAMMPS
developed by Sandia National Laboratories \cite{LAMMPS}.

\end{acknowledgments}

\bibliographystyle{unsrt}

\end{document}